# Mathematical Analysis of the Historical Economic Growth

Ron W Nielsen[1]

Environmental Futures Research Institute, Gold Coast Campus, Griffith University, Qld, 4222, Australia

**Abstract.** Data describing historical economic growth are analysed. Included in the analysis is the world and regional economic growth. The analysis demonstrates that historical economic growth had natural tendency to follow hyperbolic distributions. Parameters describing hyperbolic distributions have been determined. A search for takeoffs from stagnation to growth produced negative results. This analysis throws a new light on the interpretation of the mechanism of the historical economic growth and suggests new lines of research.

## Introduction

The latest publication of excellent data by the world-renown economist (Maddison, 2001, 2010) offers an unprecedented opportunity to study the mechanism of the historical economic growth. Earlier study (Nielsen, 2014), based on these data, indicated that historical economic growth can be described using hyperbolic distributions in much the same way as the growth of human population (von Foerster, Mora & Amiot, 1960). Unlike exponential growth, which is more familiar and which can be easier to understand, hyperbolic distributions are strongly deceptive because they appear to be made of two distinctly different components, slow and fast, joined perhaps by a certain transition component. This illusion is so strong that even the most experienced researchers can be easily deceived particularly if their research is based on a limited body of data, as it was in the past. Fortunately, Maddison's data solve this problem, and fortunately also their analysis is trivially simple because, as pointed out earlier (Nielsen, 2014), hyperbolic distributions can be easily identified and analysed using the reciprocal values of data.

Hyperbolic distribution describing *growth* is represented by a *reciprocal* of a linear function:

$$S(t) = \frac{1}{a - kt}, \tag{1}$$

---

where $S(t)$ is the size of the growing entity, in our case the Gross Domestic Product (GDP), while *a* and *k* are *positive* constants.

The reciprocal of such hyperbolic growth, $1/S(t)$, is represented by a *decreasing linear* function:

$$\frac{1}{S(t)} = a - kt. \qquad (2)$$

Hyperbolic *distributions* should not be confused with hyperbolic *functions* ($\sinh(t)$, $\cosh(t)$, etc). Furthermore, *reciprocal* functions should not be confused with *inverse* functions. Thus, for instance, for the expression given by the eqn (1) the objective of finding the inverse function would be to calculate time *t* for a given size $S(t)$. The roles of the dependent and independent variables would be reversed. For the reciprocal function, the objective is to convert eqn (1) into eqn (2). The roles of dependent and independent variables are not changed.

Reciprocal values help in an easy and generally unique identification of hyperbolic growth because in this representation hyperbolic growth is given by a decreasing straight line. Apart from serving as an alternative way to analyse data, reciprocal values allow also for the investigation of even small deviations from hyperbolic distributions because deviations from a straight line can be easily noticed.

Reciprocal values allow also for an easy identification of different components of growth. This property can be used in comparing empirical information with theoretical interpretations (Galor, 2005, 2011), which are based on the assumption of the existence of different components of growth.

When comparing mathematically-calculated distributions with the reciprocal values of data, we have to remember that the sensitivity of the reciprocal values to small deviations increases with the decreasing size *S* of the growing entity.

Suppose we have two values of *S* at a given time: $S_1$ and $S_2$, representing, for instance, the empirical and calculated values. It is clear that

$$\Delta\left(\frac{1}{S}\right) = -\frac{\Delta S}{S_1 S_2}, \qquad (3)$$

where $\Delta(1/S)$ is the difference between two inverse values and $\Delta S$ is the difference between *S* values.

For a given $|\Delta S|$, $|\Delta(1/S)|$ increases rapidly with the decreasing $S_1$ and $S_2$ values. The separation of small values of data from calculated distributions are magnified. Similar magnifications, though less pronounced, are also shown in the semilogarithmic displays of data. We shall use both displays to examine the quality of fits to the data.

It should be noted that the *decreasing* reciprocal values describe *growth*, while a deviation to *larger* reciprocal values describes decline. Consequently, a diversion to a faster trajectory will be indicated by a downward bending of a trajectory of the reciprocal values, away from an earlier observed trajectory, while the diversion to a slower trajectory will be indicated by an upward bending.

The data describing the historical economic growth (Maddison, 2001, 2010) do not allow for a detailed analysis below AD 1500 because there are two large gaps in the data: between AD 1 and 1000 and between AD 1000 and 1500. The best sets of data are from AD 1500. However, the compilation prepared by Magnuson appears to be the best and the most reliable source of data describing the historical economic growth.

Throughout the analysis presented here, the values of the Gross Domestic Product (GDP) will be expressed in billions of the 1990 International Geary-Khamis dollars. All diagrams are presented in the Appendix

Theories play an important role in scientific research because they crystallise interpretations of studied phenomena. However, theories have to be always tested by data. In science it is important to look for data confirming theoretical explanations but it is even more important to discover contradicting evidence, because data confirming a theory confirm only what we already know but contradicting evidence may lead to new discoveries.

Currently, the most complete theory describing the mechanism of the historical economic growth appears to be the Unified Growth Theory (Galor, 2005, 2008, 2011, 2012). One of the fundamental postulates of this theory is the postulate of the existence of three regimes of growth governed by three distinctly different mechanisms: (1) the Malthusian regime of stagnation, (2) the post-Malthusian regime, and (3) the sustained-growth regime.

According to Galor (2005, 2008, 2011, 2012), Malthusian regime of stagnation was between 100,000 BC and AD 1750 for developed regions and between 100,000 BC and AD 1900 for less-developed regions. The claimed starting time appears to be based entirely on conjecture because Maddison's data are terminated at AD 1 and even they contain significant gaps below AD 1500. The post-Malthusian regime was allegedly between AD 1750 and 1850 for developed regionsand and from 1900 for less-developed regions. The sustained-growth regime was supposed to have commenced around 1850 for developed regions.

Unified Growth Theory (Galor, 2005, 2008, 2011, 2012) can be tested in many ways but the easiest way to test it is to look for the dramatic takeoffs from stagnation to growth. These takeoffs are described as a "remarkable" or "stunning" escape from the Malthusian trap (Galor, 2005, pp. 177, 220). It is a signature, which cannot be missed.

This change in the pattern of growth is described as "the sudden take-off from stagnation to growth" (Galor, 2005, pp. 177, 220, 277) or as a "sudden spurt" (Galor, 2005, 177, 220). According to Galor, for developed regions, the end of the Malthusian regime of stagnation coincides with the Industrial Revolution. "The take-off of developed regions from the Malthusian Regime was associated with the Industrial Revolution" (Galor, 2005, p. 185). Indeed, the Industrial Revolution is considered to have been "the prime engine of economic growth" (Galor, 2005, p. 212).

This signature is characterised by three features: (1) it should be a prominent change in the pattern of growth, (2) it should be a transition from stagnation to growth and (3) it should occur at the time predicted by the theory. For developed regions, the postulated takeoffs should occur around AD 1750, or around the time of the Industrial Revolution, 1760-1840 (Floud & McCloskey, 1994). For less-developed regions, they should occur around 1900. The added advantage of using this simple test is that there are no significant gaps in the data around the time of the postulated takeoffs and consequently the stagnation and the expected prominent transitions from stagnation to growth should be easily identifiable.

A transition from growth to growth is not a signature of the postulated takeoff from stagnation to growth. Thus, a transition is from hyperbolic growth to another hyperbolic growth or to some other steadily-increasing trajectory is not a signature of the sudden takeoff from stagnation to growth. Likewise, a transition at a distinctly different time is not a confirmation of the theoretical expectations.

## World economic growth

Results of mathematical analysis of the world economic growth are presented in Figures 1-3. Reciprocal values of historical data can be fitted using a straight line (representing hyperbolic growth) between AD 1000 and 1955. From around 1955, the world economic growth started to be diverted to a slower trajectory as indicated by the *upward* bending of the reciprocal values. This section is magnified in Figure 2. Global economic growth is now approximately exponential (Nielsen, 2014, 2015a).

Hyperbolic fit to the world GDP data (Maddison, 2010) is shown in Figure 3. The fit is remarkably good. The point at AD 1 is 77% away from the fitted curve. We would need more data between AD 1 and 1000 to decide whether such a difference is of any significance but it could reflect a pattern similar to the pattern observed for the growth of human population (Nielsen, 2016). Hyperbolic economic growth of the historical GDP has been uniquely identified by the straight-line fitting the reciprocal values of data.

Parameters describing hyperbolic trajectory fitting the data between AD 1000 and 1955 are: $a = 1.684 \times 10^{-2}$ and $k = 8.539 \times 10^{-6}$. Its singularity is at $t = 1972$. However, from around 1955, the world economic growth started to be diverted to a slower trajectory bypassing the singularity by 17 years (see Table 1).

The search for a takeoff in the world economic growth produced negative results. The data reveal a different pattern of growth than claimed by the Unified Growth Theory (Galor, 2005, 2008, 2011, 2012). The theory claims a long period of stagnation followed by a sudden takeoff. The data show a stable hyperbolic growth followed by a diversion to a slower trajectory.

The data also demonstrate that the Industrial Revolution had no impact on changing the economic growth trajectory. These results might not be surprising because the world economic growth is represented by the economic growth in developed and less-developed regions. However, even then, it would be hard to expect that the data would follow such a remarkably stable and specific trajectory. We would expect some distortions reflecting takeoffs around the time of the Industrial Revolution for developed regions and takeoffs around 1900 for less-developed regions. We see no signs of such distortions and no signs of the presence of such takeoffs.

The straight-line representing the reciprocal values of the GDP data shown in Figure 1 follows the data closely until 1955. There was no boosting in the economic growth, no unusual acceleration at *any time* between AD 1000 and 1955. The world economic growth was increasing monotonically before and after the Industrial Revolution as shown by either a steadily increasing hyperbolic distribution in Figure 3 or by the steadily-decreasing straight line (representing hyperbolic distribution) shown in Figure 1. Which point on a straight line should be selected to mark a boundary between different patterns of growth? How can we claim different patterns of growth on a straight line if the straight line shows clearly only one pattern? There was no takeoff in the world economic growth at any time, let alone around the time of the Industrial Revolution or around 1900.

Economic growth may have been slow over a long time but it was not stagnant. The growth was hyperbolic, and the characteristic feature of hyperbolic growth is a slow growth over a long time and a fast growth over a short time. Hyperbolic growth increases monotonically and it is *impossible* to locate a place marking a transition from a slow to fast growth because *such a transitions does not exist*.

Hyperbolic growth of the world economy is in harmony with the hyperbolic growth of the world population (Nielsen, 2016; von Foerster, Mora & Amiot, 1960). In both cases, the growth was indeed slow over a long time and fast over a short time. In both cases the growth creates an illusion of stagnation followed by a sudden takeoff. However, in both cases the growth was hyperbolic. There was no stagnation and no sudden takeoff. Furthermore, in both cases the growth started to be diverted, relatively recently, to slower trajectories.

## Western Europe

The growth of the GDP in Western Europe is shown in Figures 4-6. Western Europe is represented by the total of 30 countries: Austria, Belgium, Denmark, Finland, France, Germany, Italy, the Netherlands, Norway, Sweden, Switzerland, the United Kingdom, Greece, Portugal, Spain and by 14 small, but unspecified countries. Ireland is missing in this list because it was included only from 1921.

The best hyperbolic fit to the data is between AD 1500 and 1900. Parameters for this distribution are $a = 9.859 \times 10^{-2}$ and $k = 5.112 \times 10^{-5}$. The point of singularity is at $t = 1929$. Between 1900 and 1910, economic growth started to be diverted to a slower, but still fast-increasing, trajectory bypassing the singularity by 29 years (see Table 1).

The most complete set of data for Western Europe is for Denmark, France, the Netherlands and Sweden. They are analysed separately and results are presented in Figures 7 and 8. According to Maddison (2010), these four countries accounted for 34% of the total GDP of the 30 countries of Western Europe in 2008.

Parameters describing the historical hyperbolic growth of the GDP in these four countries are: $a = 3.821 \times 10^{-1}$ and $k = 1.986 \times 10^{-4}$. The point of singularity is at $t = 1923$. From around 1875 economic growth in Denmark, France, the Netherlands and Sweden was diverted to a slower trajectory, bypassing the singularity by 48 years.

The quality of the hyperbolic fit to the data is virtually the same as for the total of the 30 countries but now the fitted curve passes also through the AD 1 point. However, it still does not reproduce the point at AD 1000. This point is only 41% below the fitted hyperbolic distribution.

The historical growth of the GDP in Western Europe was definitely hyperbolic from AD 1500 to 1900 but there is also a good indication that it might have been hyperbolic from AD 1 (see Figures 7 and 8). Even if we make allowance for this uncertainty, the search for a sudden takeoff around the expected time, i.e. around the time of the Industrial Revolution, produced negative results for the 30 countries of Western Europe and for the four (Denmark, France, the Netherlands and Sweden) characterised by the most complete sets of data.

The claim of a stunning or remarkable takeoff is contradicted by data. There was no takeoff of any kind and at any time, stunning or less stunning, remarkable or less remarkable, sudden or gradual – none at all. The Industrial Revolution, the alleged "prime engine of economic growth" (Galor, 2005a, p. 212), made no impression on changing the economic growth trajectory in regions where this engine should have been working most efficiently. Industrial Revolution brought many other important changes but, surprisingly perhaps, did not change the economic growth trajectory in the countries closest to this monumental development.

## Eastern Europe

Systematic data for Eastern Europe are available only for seven countries: Albania, Bulgaria, Czechoslovakia, Hungry, Poland, Rumania and Yugoslavia. For other countries there are no data until 1990. The analysis of the historical data for Eastern Europe is summarised in Figures 9-11.

The best hyperbolic fit to the data is between AD 1000 and 1890. Hyperbolic parameters are: $a = 7.749 \times 10^{-1}$ and $k = 4.048 \times 10^{-4}$. The point of singularity is at $t = 1915$. From around 1890, economic growth in Eastern Europe was diverted to a slower trajectory, bypassing the singularity by 25 years.

There was no stagnation and no takeoff at any time. Industrial Revolution had no impact on changing the economic growth trajectory in the countries of Eastern Europe.

## Former USSR

The analysis of the data for the countries of the former USSR is presented in Figures 12-14. The hyperbolic fit to the data is between AD 1 and 1870. Parameters fitting the data are: $a = 6.547 \times 10^{-1}$ and $k = 3.452 \times 10^{-4}$. The point of singularity is at $t = 1897$. From around 1870, or maybe even a little earlier (shortly after the Industrial Revolution) economic growth in the Former USSR was diverted to a slower trajectory, bypassing the singularity by at least 27 years.

There was no stagnation and no takeoff *at any time*. Industrial Revolution had no impact on changing the economic growth trajectory in the countries of former USSR.

## Asia

Analysis of the historical economic growth in Asia (including Japan) is summarised in Figures 15-17. The best hyperbolic fit is between AD 1000 and 1950. Parameters fitting the data are: $a = 2.303 \times 10^{-2}$ and $k = 1.129 \times 10^{-5}$. The point of singularity is at $t = 2040$.

Asia is made primarily of less-developed countries (BBC, 2014, Pereira, 2011) and consequently, according to the Unified Growth Theory (Galor, 2005, 2008, 2011, 2012), economic growth in this region should have been characterised by stagnation until around 1900, the year marking the alleged stunning escape from the Malthusian trap, the escape, which was supposed to have been manifested by the postulated dramatic takeoff. (Until AD 1900, Japan's conrtibution to the total economy in Asia was on average only 5%.) The data and their analysis show that there was no stagnation, at least from AD 1000 and no expected takeoff. The data reveal a steadily increasing hyperbolic growth until around 1950. From around that year economic growth *was* diverted to a faster trajectory. This boosting can be seen clearly in Figures 16 and 17 and it occurred close to the time of the postulated takeoff from stagnation to growth. However, it was *not* a transition from stagnation to growth but from hyperbolic growth to a slightly faster trajectory of a different kind. It is, therefore, not the takeoff postulated in the Unified Growth Theory. Furthermore, it was only a temporary boosting, which is now returning to the original hyperbolic trajectory and, as indicated by the reciprocal values of the data, this new growth is likely to be slower than the original trajectory. Thus, it is a boosting of a completely different kind. It would be interesting to explain it but we cannot be helped by Unified Growth Theory because it discusses mechanisms, which are repeatedly contradicted by data. This transition is not even recognised in this theory

Reciprocal values of data presented in Figure 16 show that the economic growth became temporarily *slower* at the time overlapping the time of the Industrial Revolution, 1760-1840 (Floud & McCloskey, 1994), because while the point in 1820 is still located on the straight line, representing hyperbolic growth, the point in 1870 is above this line. The deceleration in the economic growth occurred sometime between 1820 and 1870.

This brief deceleration was followed by a transient growth between 1870 and 1940, which appears to have been also hyperbolic but a little faster than the earlier hyperbolic growth. This transition occurred earlier than the postulated takeoff around 1900 and it was not a transition from stagnation to growth but a transition from hyperbolic growth to hyperbolic growth. Furthermore, it was also a minor transition, which could be hardly noticed in the direct display of data shown in Figure 17. In summary, therefore, the examination of data for the economic growth in Asia demonstrates that the postulated takeoff (Galor, 2005, 2008, 2011, 2012) never happened. There was no stagnation and no sudden dramatic escape to a new and rapid growth.

## Africa

Results of the analysis of the economic growth in the 57 African countries are presented in Figures 18-20. Reciprocal values of the GDP data, presented in Figures 18 and 19, show clearly that the economic growth was following *two* hyperbolic distributions. At first it was a slow hyperbolic growth between AD 1 and 1820 characterised by parameters $a = 1.244 \times 10^{-1}$ and $k = 5.030 \times 10^{-5}$ and by the singularity at $t = 2473$. Then, around 1820, this slow hyperbolic growth was replaced by a significantly faster hyperbolic growth characterised by parameters $a = 4.192 \times 10^{-1}$ and $k = 2.126 \times 10^{-4}$ and by the singularity at $t = 1972$. Defined by the parameter *k*, this new growth was 4.2 times faster than the earlier hyperbolic growth. From around 1950, this fast hyperbolic growth was diverted to a slower, non-hyperbolic trajectory, bypassing singularity by 22 years.

Africa is also made of less-developed countries (BBC, 2014; Pereira, 2011) so according to the Unified Growth Theory (Galor, 2005, 2008, 2011, 2012) it should have experienced stagnation in the economic growth until around 1900 followed by a clear takeoff around that year. These expectations are contradicted by the economic growth data because (1) economic growth was not stagnant but hyperbolic until 1950, (2) there was no takeoff from stagnation to growth around 1900 or around any other time and (3) shortly after the expected time of the takeoff, economic growth in Africa started to be diverted to a slower trajectory.

Acceleration in the economic growth in Africa occurred around 1820, but it was not a transition from stagnation to growth but *from growth to growth*. Even more specifically, it was a transition from the hyperbolic growth to another hyperbolic growth. It was also acceleration at a wrong time, not around 1900 but around the time of the Industrial Revolution. This acceleration can be explained by noticing that it appears to coincide with the intensified colonisation of Africa (Duignan & Gunn, 1973; McKay, et al. 2012; Pakenham, 1992). The fast increasing GDP after 1820 was not reflecting the rapidly improving living conditions of African population brought about by the beneficial changes caused by the Industrial Revolution but the rapidly increasing wealth of new settlers and their countries of origin at the expense of the deploring living conditions of native populations.

The search for the takeoff from stagnation to growth, claimed by the Unified Growth Theory (Galor, 2005, 2008, 2011, 2012), produced negative results. The data show also that there was no stagnation in the economic growth over the entire range of time, from AD 1 to the present time.

## Latin America

Results of the analysis of the economic growth in Latin America are presented in Figures 21 - 23. Data for Latin America are difficult to analyse because there was a significant decline in the economic growth between AD 1500 and 1600 but they also appear to follow two distinctly different hyperbolic trajectories. However, the identification of the first trajectory is not as clear as for Africa. The identification of the second hyperbolic trajectory is more convincing. Our tentative conclusion is that the economic growth in Latin America was following a slow hyperbolic distribution between AD 1 and 1500 and a fast hyperbolic distribution between AD 1600 and around 1870.

The tentatively assigned slow hyperbolic growth between AD 1 and 1500 is characterised by parameters $a = 4.421 \times 10^{-1}$ and $k = 2.093 \times 10^{-4}$. Its singularity is at $t = 2113$. The better determined fast hyperbolic growth between AD 1600 and 1870 is characterised by parameters $a = 1.570 \times 10^{0}$ and $k = 8.224 \times 10^{-4}$. Its singularity is at $t = 1910$. Defined by the parameter *k*, this growth was 3.9 times faster than the earlier hyperbolic growth. From around 1870, this fast hyperbolic growth started to be diverted to a slower trajectory bypassing the singularity by 40 years. The transition from the earlier apparent hyperbolic growth to a new and rapid hyperbolic growth, which occurred between around AD 1500 and 1600 appears to coincide with commencement of the Spanish conquest (Teeple, 2002).

Latin America is also made of less-developed countries (BBC, 2014; Pereira, 2011) so again, according to the Unified Growth Theory (Galor, 2005, 2008, 2011, 2012), the economic growth in this regions should have been stagnant until around 1900 and fast-increasing from around that year. This pattern of growth is not confirmed by data. The data show a diametrically different pattern: (1) there is no convincing evidence of the existence of stagnation over the entire range of time between AD 1 and 1870 but there is a sufficiently convincing indication of the hyperbolic growth particularly between AD 1600 and 1870, (2) there was no takeoff from stagnation to growth at any time, and (3) around the time of the postulated takeoff in 1900 there was a diversion to a slower trajectory in 1870.

Even if the identification of the hyperbolic growth between AD 1 and 1500 is questioned, the overall pattern of growth in Latin America is similar to the pattern in Africa: a slow hyperbolic growth is followed by a fast hyperbolic growth. However, in any case, there is no convincing evidence that the growth was ever stagnant. On the contrary, there is sufficiently convincing evidence that the growth was never stagnant. It was clearly not stagnant between AD 1600 and 1870.

There was also no takeoff, dramatic or modest, from stagnation to growth around the expected time of 1900, *first* because the growth before that year was not stagnant but hyperbolic and *second* because around the time of the expected remarkable takeoff the economic growth started to be diverted to a slower trajectory. The search for the postulated takeoff produced negative results.

## Summary and conclusions

Results of mathematical analysis of the historical economic growth are presented in Table 1. The listed parameters *a* and *k* are for the fitted hyperbolic distributions. The last column shows the results of the search for the takeoffs from stagnation to growth claimed by the Unified Growth Theory (Galor, 2005, 2008, 2011, 2012).

**Table 1.** *Summary of the mathematical analysis or the historical economic growth*

| Region/Countries | *a* | *k* | Hyperbolic Range | Singularity | Proximity | Takeoff |
|---|---|---|---|---|---|---|
| World | $1.684 \times 10^{-2}$ | $8.539 \times 10^{-6}$ | 1000 – 1955 | 1972 | 17 | X |
| Western Europe | $9.859 \times 10^{-2}$ | $5.112 \times 10^{-5}$ | 1500 – 1900 | 1929 | 29 | X |
| Western Europe (4) | $3.821 \times 10^{-1}$ | $1.986 \times 10^{-4}$ | 1 – 1875 | 1923 | 48 | X |
| Eastern Europe | $7.749 \times 10^{-1}$ | $4.048 \times 10^{-4}$ | 1000 – 1890 | 1915 | 25 | X |
| Former USSR | $6.547 \times 10^{-1}$ | $3.452 \times 10^{-4}$ | 1 – 1870 | 1897 | 27 | X |
| Asia | $2.303 \times 10^{-2}$ | $1.129 \times 10^{-5}$ | 1000 – 1950 | 2040 | 90 | X |
| Africa | $1.244 \times 10^{-1}$ | $5.030 \times 10^{-5}$ | 1 – 1820 | 2473 | | |
| | $4.192 \times 10^{-1}$ | $2.126 \times 10^{-4}$ | 1820 – 1950 | 1972 | 22 | X |
| Latin America | $4.421 \times 10^{-1}$ | $2.093 \times 10^{-4}$ | 1 – 1500 | 2113 | | |
| | $1.570 \times 10^{0}$ | $8.224 \times 10^{-4}$ | 1600 – 1870 | 1910 | 40 | X |

**Notes:** *a* and *k* – Hyperbolic growth parameters [see eqn (1)]. *Hyperbolic Range* - The empirically-confirmed range of time when the economic growth can be described using hyperbolic distributions. *Singularity* - The time of the escape to infinity for a given hyperbolic distribution. *Proximity* - Proximity (in years) of the singularity at the time when the economic growth departed from the hyperbolic growth to a new trajectory. *Western Europe (4)* - Four countries of Western Europe: Denmark, France, the Netherlands and Sweden. *X* - No takeoff. The takeoff from stagnation to growth claimed by the Unified Growth Theory (Galor, 2005, 2008, 2011, 2012) never happened.

This analysis demonstrates that the natural tendency for the historical economic growth was to increase hyperbolically. In general, there is a remarkably good agreement between the data and the calculated hyperbolic distributions.

Unlike the more familiar exponential distributions, which are easier to understand because they show more readily a gradually increasing growth, hyperbolic distributions appear to be made of two or maybe even three components: a slow component, a fast component and perhaps even a transition component located between the apparent slow and fast components. This illusion is so strong that even the most experienced researchers can be deceived particularly if they have no access to good sets of data, which was in the past. Now, however, excellent data are available (Maddison, 2001, 2010) and we can use them not only to check the earlier interpretations of economic growth but also to expand the scope of the economic research.

The postulate of the existence of the epoch of Malthusian stagnation is suggested by a slow economic growth over a long time but this slow growth is just a part of the hyperbolic growth, which is convincingly identified using reciprocal values. Hyperbolic distributions create also the illusion of a sudden takeoff but this feature is also a part of the hyperbolic growth. Hyperbolic growth *is* slow over a long time and fast over a short time but the slow and fast growth are the integral features of the same monotonically increasing distribution, which is easier to understand by using the reciprocal values of the growing entity (Nielsen, 2014). In such displays, the illusion of distinctly different components disappears because hyperbolic growth is then represented by a decreasing straight line, which is easy to understand. It then becomes obvious that hyperbolic distribution cannot be divided into distinctly different sections governed by different mechanism because it makes no sense to divide a straight line into arbitrarily chosen sections and claim different mechanism to such arbitrarily-selected sections. It is also then clear that it is *impossible* to pinpoint the transition from a slow to a fast growth. Which point on a straight line should we select to identify such a transition? The transition does not happen at any specific time but gradually over the whole range of time.

Our search for the postulated takeoffs from stagnation to growth (Galor, 2005, 2008, 2011, 2012) produced negative results: *there were no takeoffs*. Galor's elaborate discussion revolving around his postulated three regimes of growth and the postulated takeoffs from stagnation to growth are irrelevant because there were no takeoffs in the growth of the GDP and in the growth of income per capita (GDP/cap) (Nielsen, 2015b). In science, just one contradicting evidence in data is sufficient to show that a theory advocating the contradicted postulate or postulates has to be either rejected or revised to bring it in the agreement with empirical evidence. In the case of the Unified Growth Theory (Galor, 2005, 2008, 2011, 2012), the postulated takeoffs from stagnation to growth are contradicted repeatedly by the economic growth in Western Europe, Eastern Europe, former USSR, Asia, Africa, and Latin America as well as by the world economic growth.

The data and their analysis suggest new lines of research of economic growth. They suggest that our attention should not be directed towards explaining the mechanism of stagnation and of the sudden takeoffs from stagnation to growth because these features are contradicted by data. What needs to be explained is why the historical economic growth was hyperbolic and why relatively recently it was diverted to a slower trajectory. Maddison published excellent data describing not only economic growth but also the growth of human population and these data can be used effectively in trying to explain the historical economic growth.

# Appendix

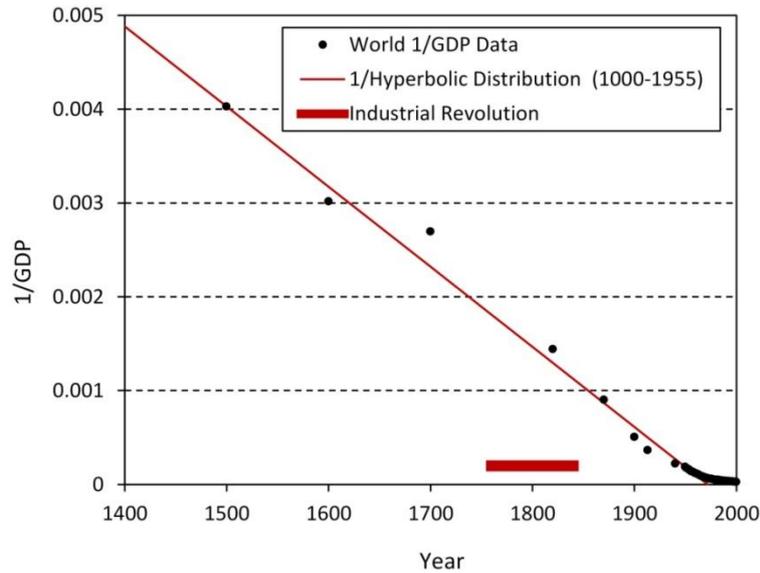

**Figure 1**. *Reciprocal values of the GDP data (Maddison, 2010) are fitted using straight line between AD 1000 and 1955 representing hyperbolic growth. There was no stagnation and no takeoff from stagnation to growth, claimed by the Unified Growth Theory (Galor, 2005, 2008, 2011, 2012). Industrial Revolution had no impact on changing the economic growth trajectory. From around 1955, the economic growth started to be diverted to a slower trajectory.*

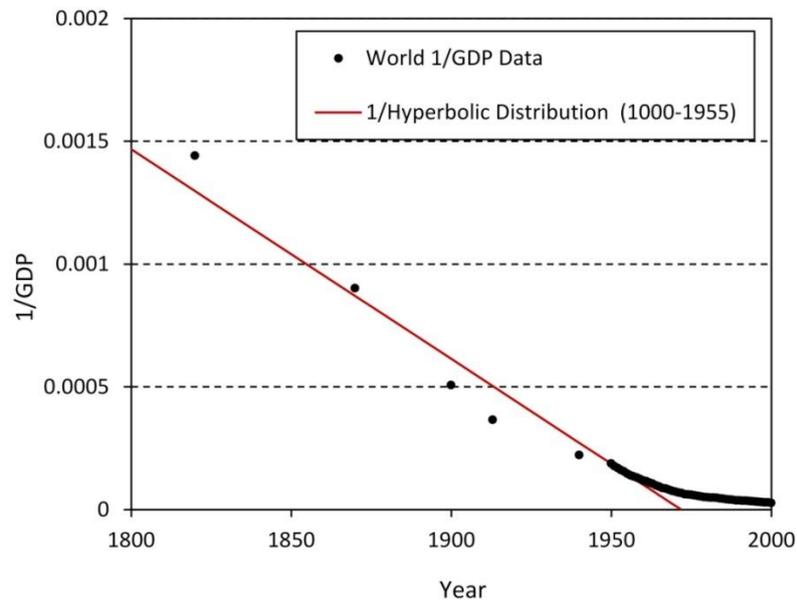

**Figure 2.** *Reciprocal values of the GDP data (Maddison, 2010) showing the diversion of the economic growth to a slower trajectory from around 1955, as indicated by the upward bending. The current global economic growth is approximately exponential (Nielsen, 2014, 2015a).*

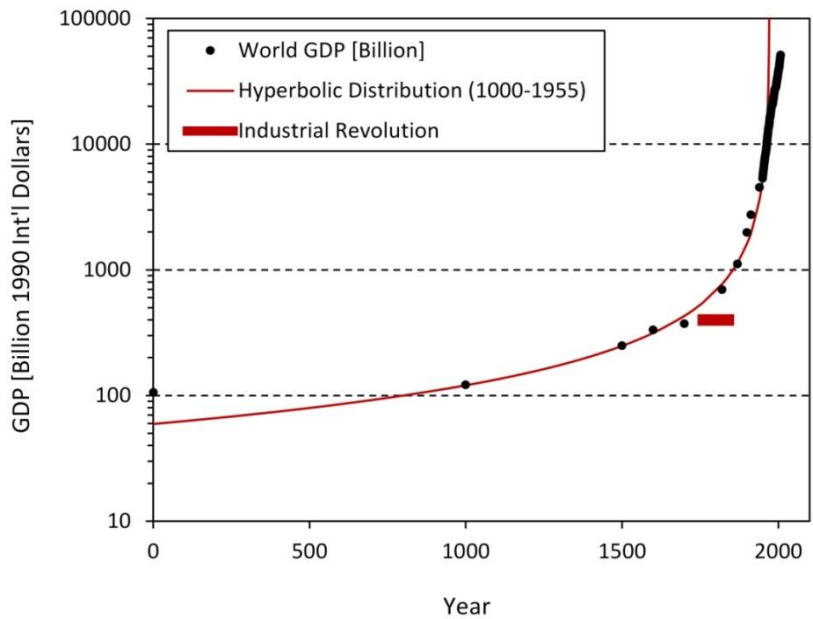

**Figure 3.** *World GDP data (Maddison, 2010) fitted using hyperbolic distribution. The point at AD 1 is 77% higher than the calculated distribution. There was no stagnation and no takeoff from stagnation to growth. Both features were incorrectly claimed by the Unified Growth Theory (Galor, 2005, 2008, 2011, 2012). Industrial Revolution had no impact on changing the economic growth trajectory. From around 1955, the world economic growth started to be diverted to a slower but still fast-increasing trajectory, which is now approximately exponential (Nielsen, 2014, 2015a).*

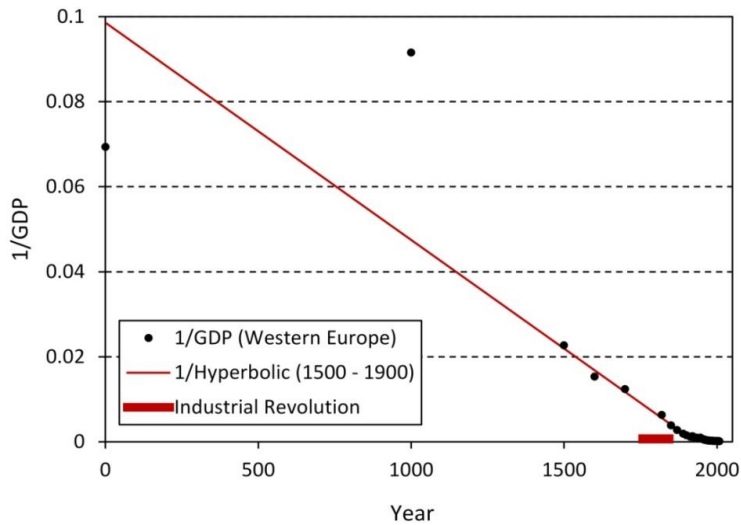

**Figure 4.** *Reciprocal values of the GDP data (Maddison, 2010) for Western Europe are compared with the hyperbolic distribution represented by the decreasing straight line. The growth was hyperbolic from at least AD 1500 to 1900. There was no takeoff from stagnation to growth. Industrial Revolution had no impact on changing the economic growth trajectory in Western Europe, the centre of this revolution. On the contrary, from around 1900, shortly after the Industrial Revolution, the economic growth in Western Europe started to be diverted to a slower trajectory as indicated by the upward bending of the trajectory representing the reciprocal values of data.*

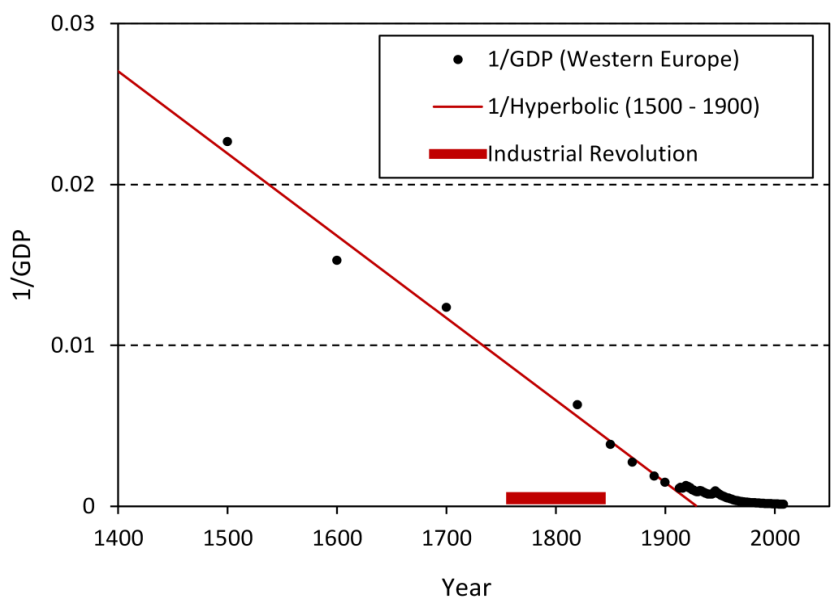

**Figure 5.** *Reciprocal values of the GDP data (Maddison, 2010) for Western Europe between AD 1500 and 2008 showing a diversion to a slower trajectory from around 1900. There was no takeoff from stagnation to growth, claimed incorrectly by the Unified Growth Theory (Galor, 2005, 2008, 2011, 2012). Industrial Revolution had absolutely no impact on changing the economic growth trajectory in Western Europe, the centre of this revolution.*

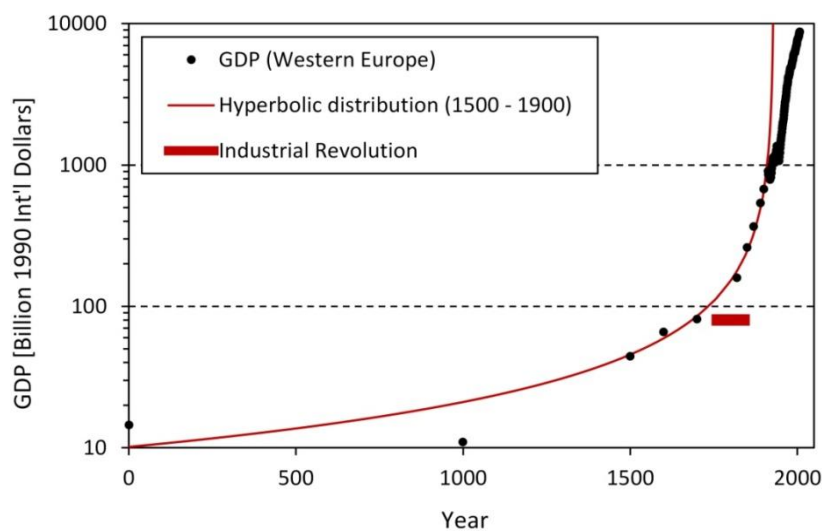

**Figure 6.** *Economic growth in Western Europe. The GDP data (Maddison, 2010) are compared with hyperbolic distribution. The growth was hyperbolic from at least AD 1500 to around 1900. The point at AD 1 is 42% higher than for the calculated distribution and 48% lower at AD 1000. There was no takeoff from stagnation to growth, claimed incorrectly by the Unified Growth Theory (Galor, 2005, 2008, 2011, 2012). Industrial Revolution had no impact on changing the economic growth trajectory in Western Europe, the centre of this revolution. From around 1900, economic growth in Western Europe started to be diverted to a slower trajectory.*

**Denmark, France, Netherlands and Sweden**

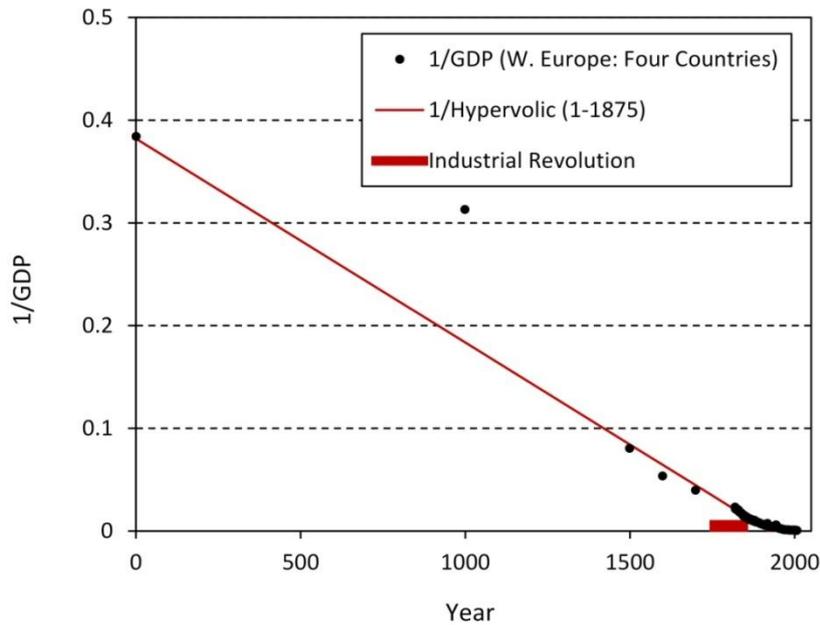

**Figure 7.** *Reciprocal values of the GDP data (Maddison, 2010) describing economic growth in four countries of Western Europe (Denmark, France, the Netherlands and Sweden) compared with the straight line representing hyperbolic growth fitting the data between AD 1 and 1875. From around 1875, or shortly after the Industrial Revolution, economic growth in these four countries started to be diverted to a slower trajectory. Industrial Revolution did not boost economic growth. There was no takeoff from stagnation to growth.*

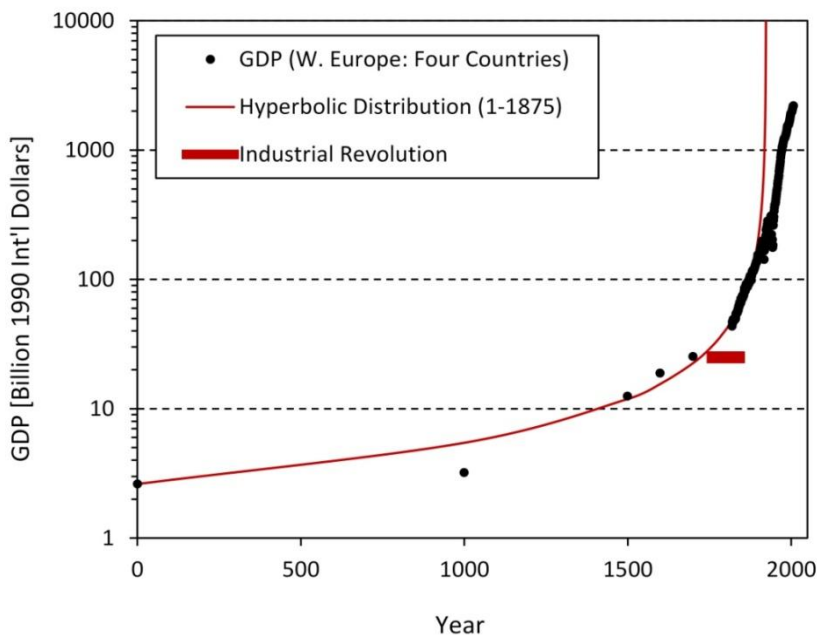

**Figure 8.** *Economic growth in Denmark, France, the Netherlands and Sweden. The data (Maddison, 2010) are compared with hyperbolic distribution. The point at AD 1000 is 41% lower than for the calculated distribution. From around 1875, the economic growth started to be diverted to a slower trajectory. There was no takeoff from stagnation to growth.*

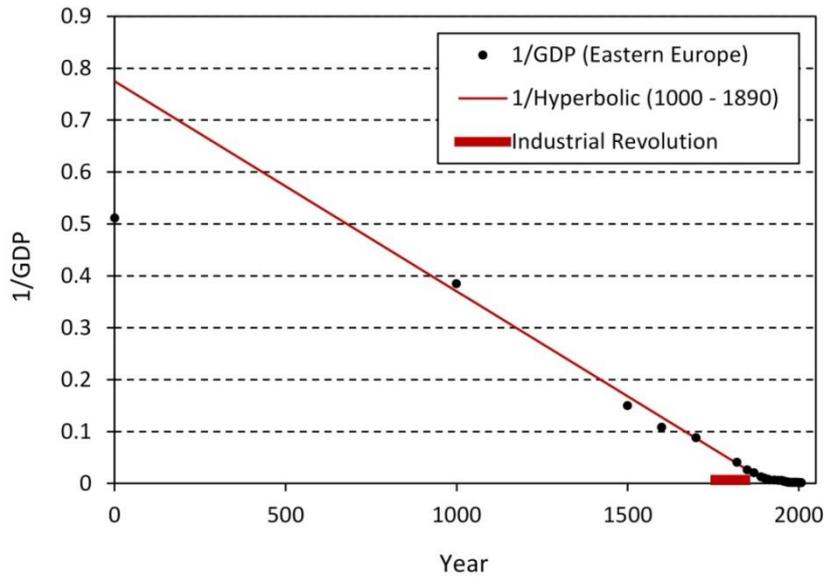

**Figure 9.** *Reciprocal values of the GDP data (Maddison, 2010) for Eastern Europe are compared with the hyperbolic distribution represented by the decreasing straight line. Economic growth was hyperbolic from at least AD 1000. The takeoff from stagnation to growth never happened because there was no stagnation. Industrial Revolution did not boost the economic growth in Eastern Europe.*

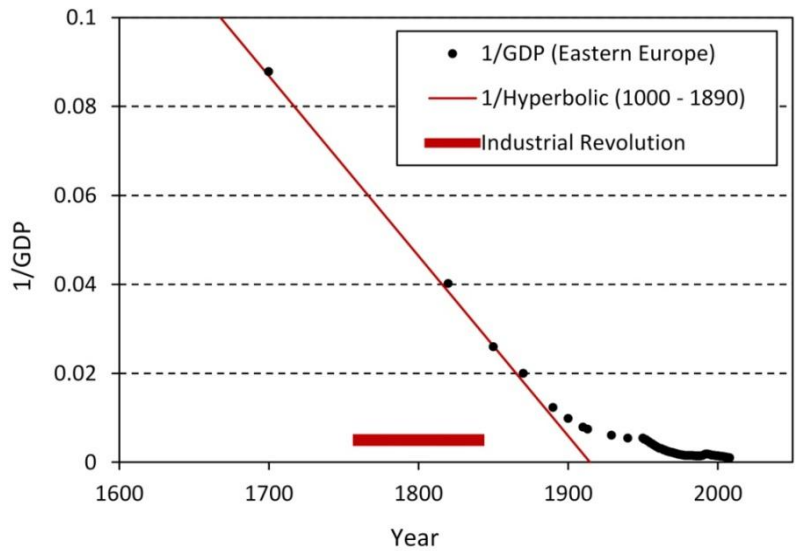

**Figure 10.** *Reciprocal values of the GDP data (Maddison, 2010) for Eastern Europe showing that from around 1890, shortly after the Industrial Revolution, the economic growth started to be diverted to a slower trajectory. There was no takeoff from stagnation to growth because there was no stagnation. Industrial Revolution did not boost the economic growth in Eastern Europe. Hyperbolic growth around that time remained undisturbed.*

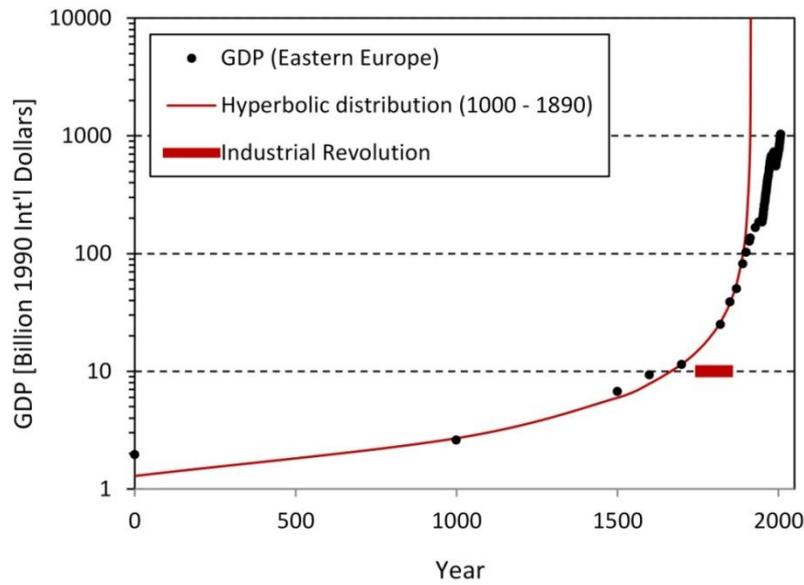

**Figure 11**. *Economic growth in Eastern Europe. GDP data (Maddison, 2010) are compared with the best hyperbolic fit. The point at AD 1 is 51% higher than for the calculated distribution. From around 1890, shortly after the Industrial Revolution, economic growth started to be diverted to a slower trajectory. Industrial Revolution did not boost the economic growth in Eastern Europe. Contrary to the Unified Growth Theory (Galor, 2005, 2008, 2011, 2012), there was no stagnation and no takeoff from stagnation to growth.*

**Former USSR**

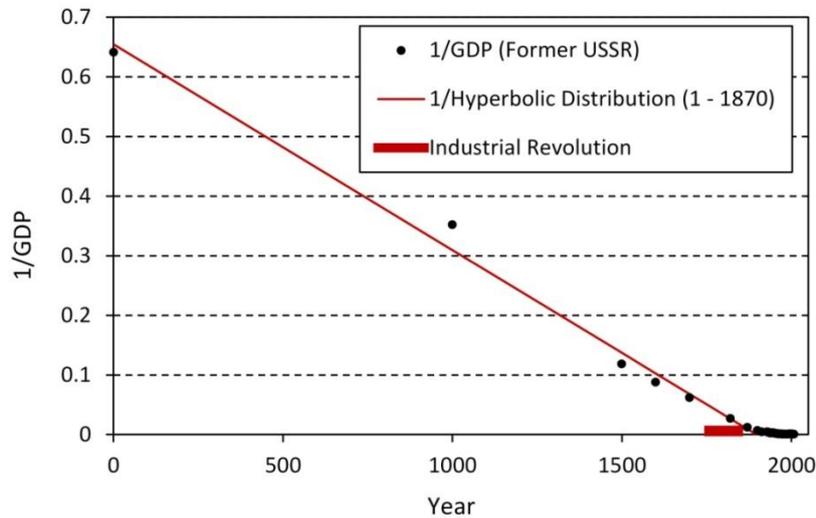

**Figure 12.** *Reciprocal values of the GDP data (Maddison, 2010) for the former USSR compared with the hyperbolic distribution represented by the decreasing straight line. Data indicate that the economic growth was hyperbolic from AD 1 to 1870. Industrial Revolution did not boost the economic growth. There was no stagnation and no takeoff from stagnation to growth. Shortly after the Industrial Revolution, the economic growth in Eastern Europe started to be diverted to a slower trajectory. Unified Growth Theory (Galor, 2005, 2008, 2011, 2012) is contradicted by the economic growth data.*

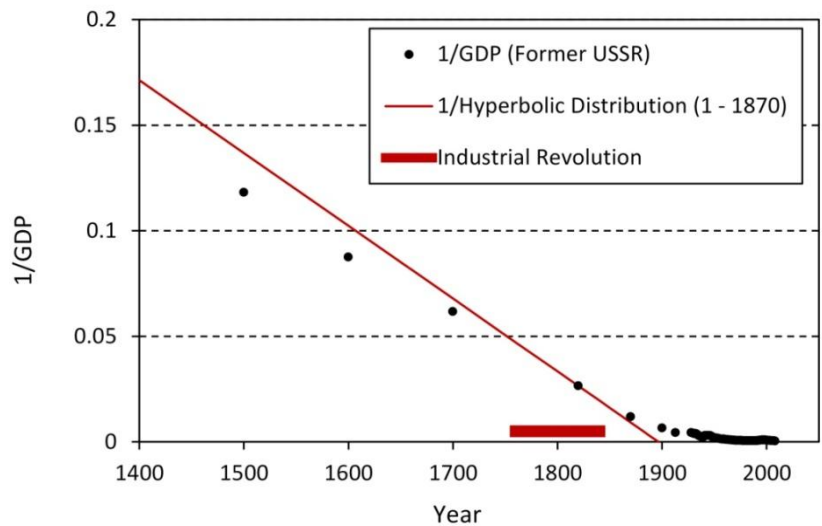

**Figure 13.** *Reciprocal values of the GDP data (Maddison, 2010) for the former USSR showing that from around 1870, shortly after the Industrial Revolution, economic growth started to be diverted to a slower trajectory.*

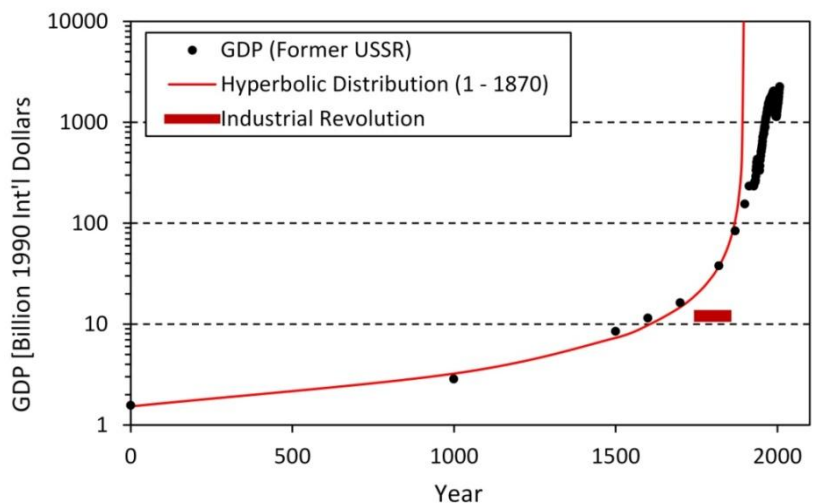

**Figure 14.** *Economic growth in the former USSR. GDP data (Maddison, 2010) are compared with the hyperbolic fit. The growth was hyperbolic from AD 1 to 1870. From around 1870, shortly after the Industrial Revolution, economic growth started to be diverted to a slower trajectory. Epoch of stagnation did not exist in the economic growth. Industrial Revolution did not boost the economic growth. There was no takeoff from stagnation to growth because there was no stagnation but a steadily-increasing growth. Unified Growth Theory (Galor, 2005, 2008, 2011, 2012) is contradicted by the economic growth data.*

**Asia (including Japan)**

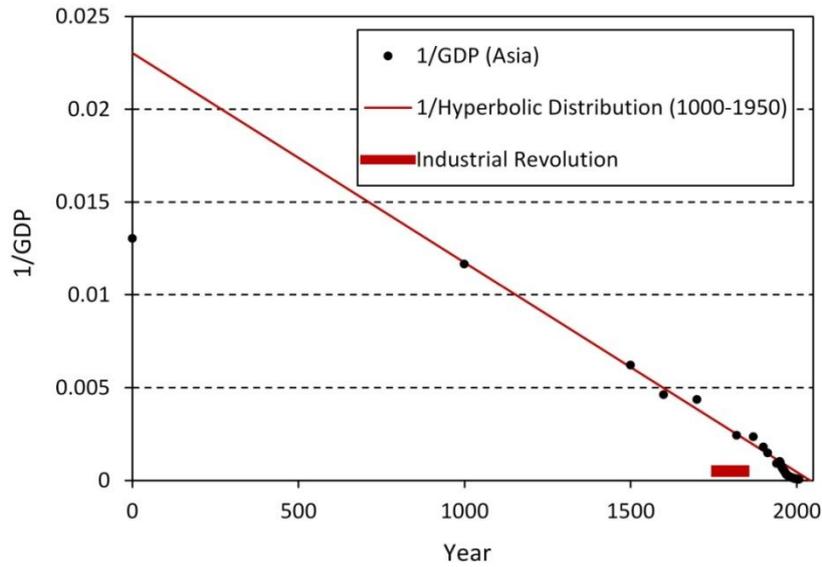

**Figure 15.** *Reciprocal values of the GDP data (Maddison, 2010) for Asia (including Japan) compared with the hyperbolic distribution represented by the decreasing straight line. Economic growth was hyperbolic from at least AD 1000. There was no expected transition from stagnation to growth.*

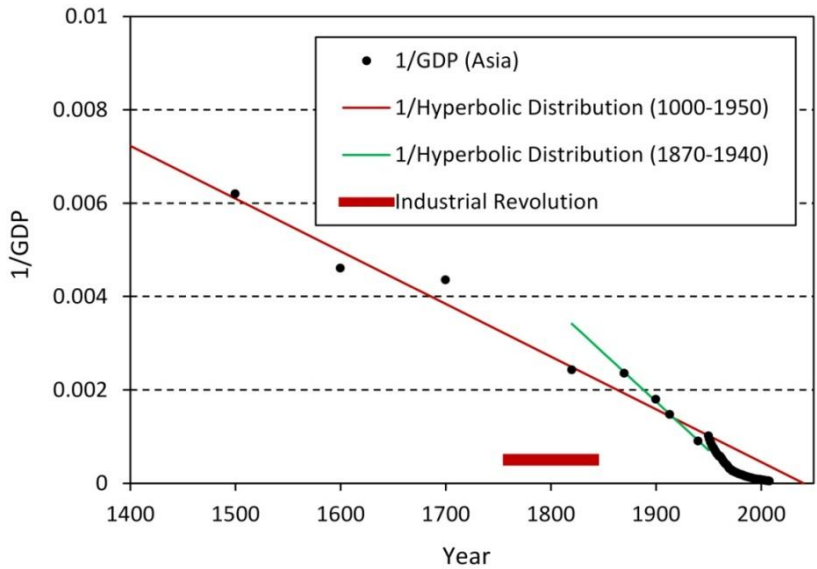

**Figure 16.** *Reciprocal values of the GDP data (Maddison, 2010) for Asia (including Japan). The data show a minor deceleration of growth towards the end of the time of the Industrial Revolution followed by a slightly faster hyperbolic growth between 1870 and 1940. The expected takeoff from stagnation to growth around 1900 (Galor, 2005, 2008, 2011, 2012) did not happen. The data show a small boosting around 1950 but it was not a transition from stagnation to growth. The search for the postulated takeoff (Galor, 2005, 2008, 2011, 2012) produced negative results.*

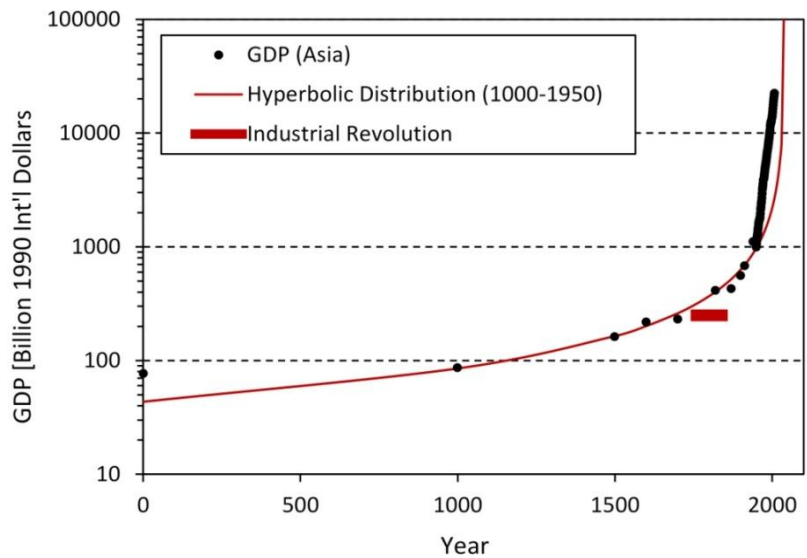

**Figure 17.** *Economic growth in Asia (including Japan). The data (Maddison, 2010) are compared with the hyperbolic distribution. The point at AD 1 is 76% higher than the calculated value. The data show a minor boosting around 1950 but it was not a transition from stagnation to growth but from the hyperbolic growth to a slightly faster trajectory, which is now coming closer to the earlier hyperbolic trajectory. The boosting was not only small but also it did not last long. The search for the postulated takeoff from stagnation to growth (Galor, 2005, 2008, 2011, 2012) produced negative results.*

**Africa**

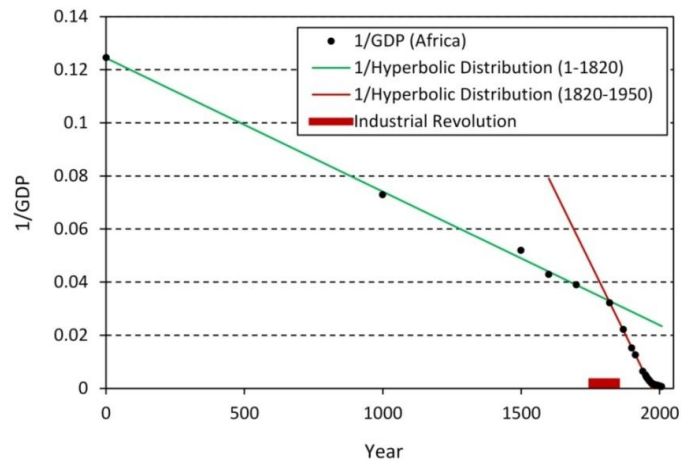

**Figure 18.** *Reciprocal values of the GDP data (Maddison, 2010) for Africa compared with hyperbolic distributions represented by the decreasing straight lines. There was no stagnation in the economic growth. Economic growth was increasing hyperbolically between AD 1 and around 1820 and again from 1820 to around 1950. The expected takeoff from stagnation to growth (Galor, 2005, 2008, 2011, 2012) never happened. The acceleration around 1820 was not a transition from stagnation to growth but from growth to growth. It also occurred earlier than expected (in 1820 rather than around 1900). Furthermore, close to the postulated takeoff in 1900, economic growth started to be diverted to a slower trajectory. The search for the takeoff from stagnation to growth around 1900 produced negative results. Unified Growth Theory (Galor, 2005, 2008, 2011, 2012) is contradicted by data.*

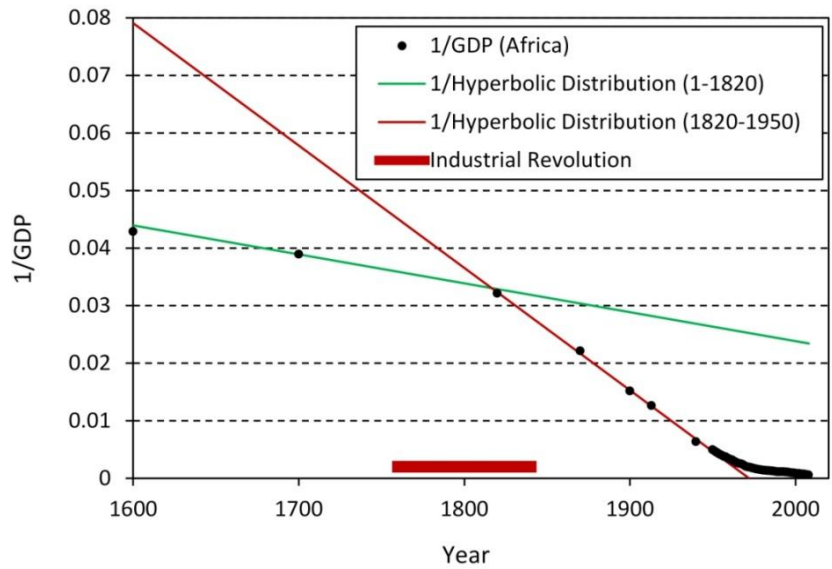

**Figure 19.** *Reciprocal values of the GDP data (Maddison, 2010) for Africa showing that from around 1950 economic growth started to be diverted to a slower trajectory. There was no takeoff around 1900, not even from growth to growth.*

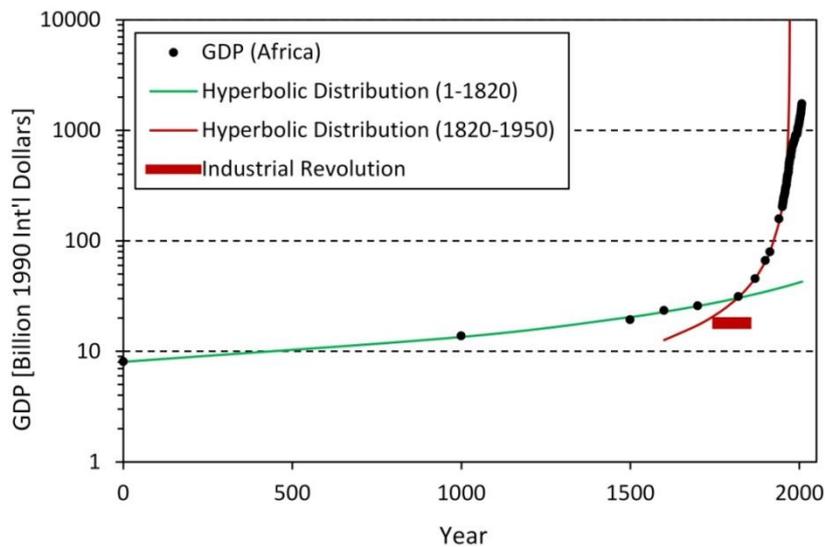

**Figure 20.** *Economic growth in Africa. Data (Maddison, 2010) are compared with hyperbolic distributions. The claimed takeoff from stagnation to growth (Galor, 2005, 2008, 2011, 2012) never happened because there was no stagnation. Furthermore, the transition from hyperbolic growth to hyperbolic growth occurred earlier (around 1820) than the postulated takeoff from stagnation to growth (around 1900). From around 1950, close to the claimed but non-existing takeoff from stagnation to growth, economic growth started to be diverted to a slower trajectory. Unified Growth Theory (Galor, 2005, 2008, 2011, 2012) is contradicted by data.*

**Latin America**

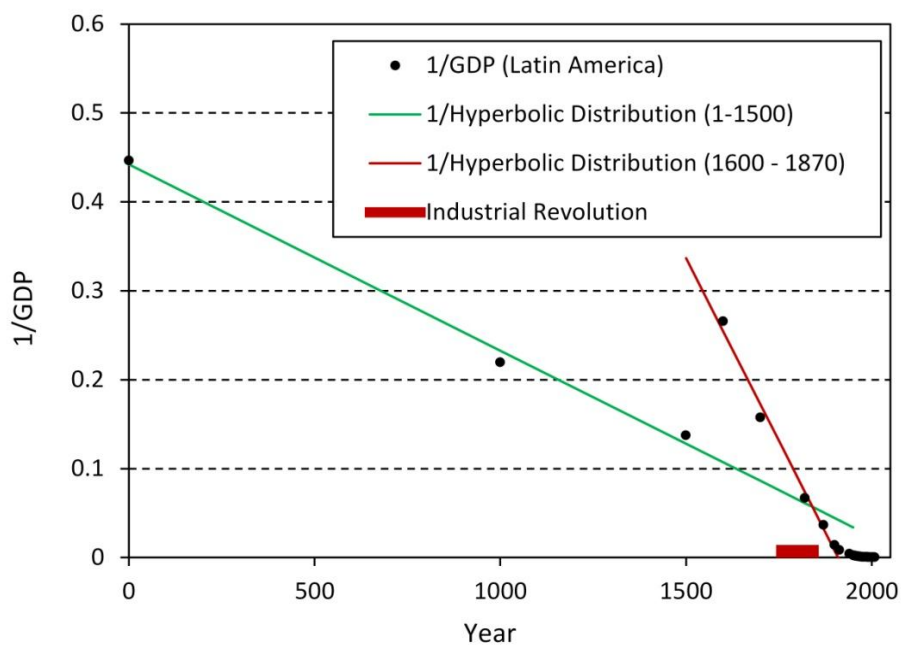

**Figure 21.** *Reciprocal values of the GDP data (Maddison, 2010) for Latin America are compared with hyperbolic distributions represented by the decreasing straight lines. The pattern of growth in Latin America is similar to the pattern of growth in Africa. The expected takeoff from stagnation to growth around 1900 (Galor, 2005, 2008, 2011, 2012) did not happen, because there was no stagnation and because, from around 1870, economic growth started to be diverted to a slower trajectory.*

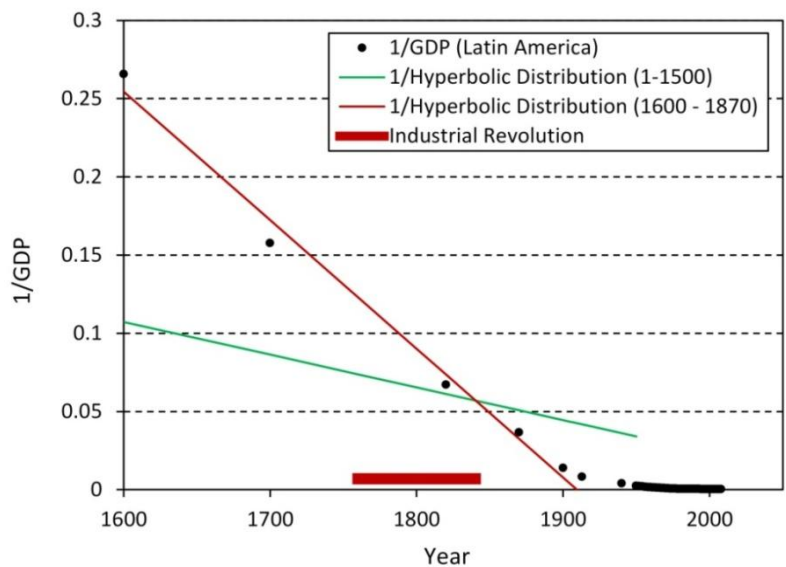

**Figure 22.** *Reciprocal values of the GDP data (Maddison, 2010) for Latin America showing that from around 1870, i.e. close to the time of the expected takeoff (around 1900) from stagnation to growth (Galor, 2005, 2008, 2011, 2012) economic growth started to be diverted to a slower trajectory. The data show also that the takeoff from stagnation to growth could not have happened because there was no stagnation.*

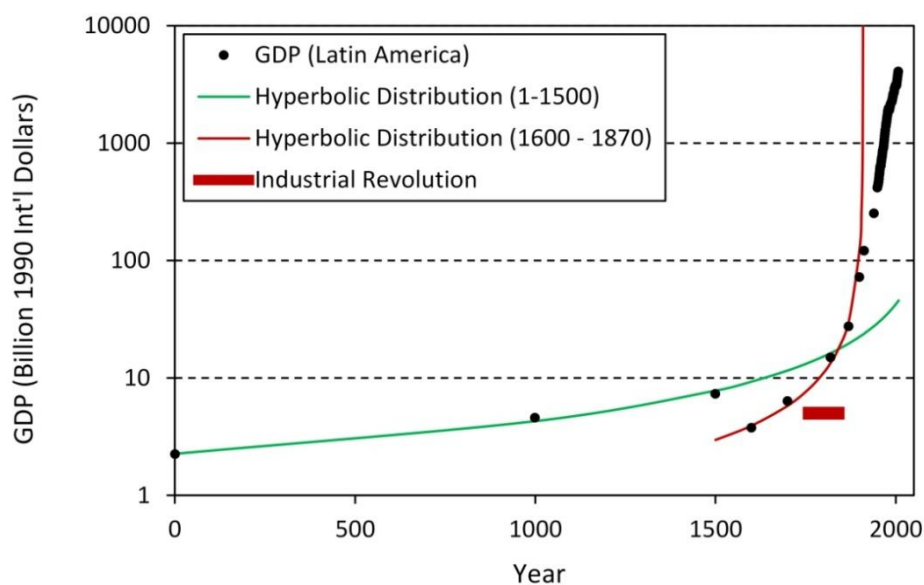

**Figure 23.** *Economic growth in Latin America. Economic growth data (Maddison, 2010) are compared with hyperbolic distributions. Unified Growth Theory (Galor, 2005, 2008, 2011, 2012) is contradicted by data. Economic growth was not stagnant before the postulated takeoff from stagnation to growth (around 1900) but hyperbolic. The growth was also stable and hyperbolic around the time of the Industrial Revolution in the Western world. The transition from stagnation to growth could not have happened because there was no stagnation. Furthermore, from around 1870, i.e. from around the time of the postulated takeoff, economic growth started to be diverted to a slower trajectory. The search for the takeoff from stagnation to growth produced negative results.*